\documentclass[12pt]{iopart}
\usepackage{iopams}
\usepackage{epsf}
\usepackage{graphicx}

\begin{document}
\submitto{\jpb}
\title[submitted version]{Polarization phase matrices for radiation 
scattering on atoms in external magnetic fields\,: \\{\it The case of 
forbidden transitions in astrophysics}} 

\author{Yee Yee Oo$^1$, Phyu Phyu San$^1$,
M. Sampoorna$^2$, K.~N. Nagendra$^2$ and G. Ramachandran$^3$}
                                                                                
\address{$^1$ Department of Physics, Mandalay University, Myanmar}
\address{$^2$ Indian Institute of Astrophysics, Bangalore 560 034, India}
\address{$^3$ GVK Academy, Jayanagar, Bangalore 560 082, India}

\begin{abstract} 
Using a quantum electrodynamical approach, we derive the scattering 
phase matrices for polarized radiation in forbidden line transitions and in the 
presence of an external magnetic fields. The case of ($J=0 \to 2 \to 0$) 
scattering is considered as an example. The non-magnetic Rayleigh scattering 
phase matrix is also presented. The Stokes profiles in a single scattering 
event are computed for the strong field (Zeeman) and weak field (Hanle) 
limits, covering also the regime of intermediate field strengths (Hanle-Zeeman). 
\end{abstract}


\section{Introduction}   
In astronomical spectroscopy, the so-called forbidden transitions are as 
important as the allowed transitions \cite{neb04}. They are seen 
\cite{neb04,bra83,ost89,shull82,chandra83,bhatia95,kastner97,judge98,lin00,
casini02,charr02,ryde06,ryde07,ryde062,aldenius07,caffau07,caffau072} 
in the spectra of rarefied gases and plasmas under special conditions, such 
as are found in some nebulae, solar corona, parts of active galactic nuclei 
and the extreme upper atmosphere of the earth. Historically, the term 
``forbidden" was associated with all those atomic transitions which do not 
involve emission or absorption of the electric dipole radiation. The forbidden 
lines may some times account for $90\%$ or more of the total visual brightness 
of an object like a planetary nebula, so much so that, when such lines were 
first seen in the $1860$'s, they were thought to be due to a new element which 
was called as Nebulium. Forbidden lines are often unaffected by departures 
from local thermodynamic equilibrium (LTE) and are also insensitive to 
atmospheric temperature uncertainties as well as micro-turbulence. As such they 
have been quite useful to measure abundances, because at low densities of the 
order of a few atoms per cubic centimeter, the strength of a spectral line 
grows as $N^{2}$ where $N$ denotes the total number of atoms in the concerned 
atomic state, whereas at high densities it grows as $N$. Scattering of 
radiation by atoms involves absorption followed by spontaneous emission and 
the scattering phase function has to be generalized~\cite{sekhar50} to a 
$4 \times 4$ phase matrix to include polarization state of the incoming or 
outgoing radiation. 

With technological advances in spectropolarimetry, 
accuracies of the order $10^{-5}$ have been achieved in measuring the 
polarization expressed through Stokes parameters. Using such instruments, it 
has been possible to discover a wealth of new and unexpected phenomena that 
are taking place in the outer layers of the solar atmosphere~\cite{stenflo04}. 
These recent observations have opened a new window to the diagnosis of the 
polarized light scattering phenomena in the Sun and the 
stars~\cite{stenflo94,stenflo96,nagendra99,trujillo02,casini05,nagendra07,
henoux,trujillo022,landi04,gandorfer00}. It is well-known that magnetic fields are
present almost everywhere in the universe~\cite{richard}. Therefore we focus 
our attention on scattering in the presence of magnetic fields. In our 
previous papers we developed a quantum electrodynamical (QED) approach to line 
scattering in the presence of external electric and magnetic fields of 
arbitrary strengths~\cite{oo1,oo2}. That approach can take account of 
all multipole type transitions, apart from the dominant one, namely, the 
dipole scattering. In this paper, we restrict our attention to 
the case of coherent scattering 
in the laboratory frame. The QED theory of dipole scattering, for the more 
realistic case of complete frequency redistribution (CRD) and 
partial frequency redistribution 
(PRD) are respectively developed by Landi Degl'Innocenti \& Landolfi 
(\cite{landi04}, and references cited therein) and by 
Bommier \cite{bom97a,bom97b,bom99a,bom03}. 
These authors consider Hanle scattering in the presence of only magnetic 
fields of arbitrary strength. The classical non-perturbative theory of 
Hanle-Zeeman scattering with PRD in atom's rest frame is presented in 
\cite{bom99}. Sampoorna et al. \cite{sam07a,sam07b} present the corresponding 
laboratory frame expressions in a form suitable for astrophysical applications. 

A classical theory for the coronal forbidden emission lines that arise from 
magnetic dipole ($M1$) transition is derived in \cite{lin00} and \cite{casini02}. 
The purpose of the present paper is mainly to derive the phase matrices
characterizing atomic scattering of polarized radiation for 
forbidden transitions in external magnetic fields. This approach can 
handle all multipoles in the transition. In particular, we 
present the explicit forms of the phase matrices for $E2$  and $M2$ type 
transitions, as they are not available in the literature. In sections 2 and 3 
we present the theoretical framework briefly. In section 4 we describe 3 limiting 
cases of scattering in magnetic fields. The numerical examples illustrating the 
application of these formulae are given in section 5. 

\section{Theory of scattering for forbidden transitions}
It is well-known that an energy level of an atom with total angular momentum
${J}$ splits into $(2J+1)$ equally spaced components (Zeeman splitting) 
when the atom is exposed to an external magnetic field ${\pmb B}$. 
These levels are characterized by states $|Jm\rangle$, where $m$ denotes the 
magnetic quantum number which is projection of the total angular momentum
operator ${\pmb J}$ along ${\pmb B}$, which is taken along the 
$Z-$axis. If we consider atomic scattering $J_{i} \to J \to J_{f}$ of 
radiation, the elements of the on-energy-shell transition matrix 
${\bf T}$ are given~\cite{oo1,oo2} by 
\begin{eqnarray}
\langle J_f m_f|{\bf T}|J_i m_i\rangle=\sum_{m}\langle J_{f}m_{f}|
{\mathcal E}({\pmb k}, 
\mu)|J m \rangle \varphi_{m} \langle J m|{\mathcal A}({\pmb k}^{\prime}, 
\mu^{\prime})|J_{i}m_{i}\rangle \ , \label{tmatrix}
\end{eqnarray}
where $\varphi_{m}$ denotes the profile function 
\begin{equation}
\varphi_{m}= (\omega_{mm_f}-\omega-{\rm i}\Gamma)^{-1} \ ; \quad 
\omega_{m m_f}=E_m - E_f \ . 
\end{equation}
Here $\omega=2\pi\nu$ denotes the frequency of the scattered radiation, 
$E_{i},\, E_{m}$ and $E_{f}$ denote respectively the energies associated with
the initial $|J_i m_i\rangle$, intermediate $|J m\rangle$ and final 
$|J_f m_f\rangle$ states of the atom, and $\Gamma$ denotes the natural width 
associated with the intermediate state. The vectors ${\pmb k}^{\prime}$ 
$(k^{\prime}, \theta^{\prime}, \phi^{\prime})$ and ${\pmb k}$ 
$(k, \theta, \phi)$ are photon momenta of the incident and scattered 
radiation. The states of circular polarization associated with 
${\pmb k}^{\prime}$ and ${\pmb k}$ are denoted respectively by 
$\mu^{\prime},\mu = \pm 1$ following the convention
employed by Rose~\cite{rose}. If ${\pmb \rho}({\pmb k}^{\prime})$ denotes
the $2 \times 2$ density matrix representing the state of polarization
of the incident radiation, the density matrix of the scattered radiation 
is given by
\begin{equation}
{\pmb \rho}({\pmb k})= {\bf T}{\pmb \rho}({\pmb k}^{\prime}){\bf T}^{\dag},
\label{rho}
\end{equation}
where ${\bf T}^{\dag}$ denotes the hermitian conjugate of ${\bf T}$ considered
as a $2 \times 2$ matrix with respect to the basis states of circular
polarization. Following~\cite{oo1,oo2,oo3,oo4}, the matrix elements for emission
and absorption processes in Eq.~(\ref{tmatrix}) are given by
\begin{equation}
\!\!\!\!\!\!\!\!\!\!\!\!\!\!
\langle J_{f}m_{f}|{\mathcal E}({\pmb k}, \mu)|J m\rangle =
\sum_{L} C(J_{f}LJ;m_{f}Mm) D^{L}_{M\mu}(\phi,\theta,0)^{*}
(-{\rm i}\mu)^{h(L)} {\mathcal J}_{L}(\omega)^{*},
\end{equation}
\begin{equation}
\!\!\!\!\!\!\!\!\!\!\!\!\!\!\!\!\!\!
\langle J m|{\mathcal A}({\pmb k}^{\prime}, \mu^{\prime})|
J_{i}m_{i}\rangle =
\sum_{L^{\prime}} C(J_{i}L^{\prime}J;m_{i}M^{\prime}m) 
D^{L^{\prime}}_{M^{\prime}\mu^{\prime}}(\phi^{\prime},\theta^{\prime},0)
(-{\rm i}\mu^{\prime})^{h(L^{\prime})} {\mathcal J}_{L^{\prime}}
(\omega^{\prime}),
\end{equation}
where $\omega=k,\ \omega^{\prime}=k^{\prime}$ and ${\mathcal J}_{L}(\omega)$
denotes the reduced matrix elements for the atomic transition from a lower level
to an upper level given by Eq.~(31) of \cite{oo1}. Further 
\begin{equation}
h(L)=\frac{1}{2}\big[ 1+\pi_{f}\pi(-1)^{L} \big]\ ;\quad 
h(L^{\prime})=\frac{1}{2}\big[ 1+\pi_{i}\pi(-1)^{L^{\prime}} \big], 
\end{equation}
with $\pi_{i},\, \pi$ and $\pi_{f}$ denoting respectively the parities of the
initial, intermediate and final atomic states. All transitions other than
$L=L^{\prime}=\pi\pi_{f}=\pi\pi_{i}=1$ are referred to as forbidden, following 
the traditional usage. 

\section{Hanle-Zeeman scattering phase matrix}
Any density matrix ${\pmb \rho}$ for polarized radiation may be expressed 
in terms of the corresponding Stokes parameters $S_p$ with $p=0,1,2,3$, through 
\begin{equation}
{\pmb \rho} = \frac{1}{2} \sum_{p=0}^{3} {\pmb \sigma}_{p} S_{p}\ , 
\end{equation}
where ${\pmb \sigma}_{0}$ denotes the $2 \times 2$ unit matrix and 
${\pmb \sigma}_{1,2,3}$ denote the well-known Pauli matrices. 
The parameters $(S_{0}, S_{1}, S_{2}, S_{3}) \equiv (I, Q, U, V)$ denote 
the Stokes vector ${\pmb S}$. The $4 \times 4$ scattering matrix 
${\pmb {\cal R}}$ is now defined through the relation 
\begin{equation}
{\pmb S}({\pmb k}) = {\pmb {\cal R}}\, {\pmb S}^{\prime}({\pmb k}^{\prime})\ .
\end{equation}
Noting that 
\begin{equation}
S_{p}=tr({\pmb \sigma}_{p}\, {\pmb \rho}) \ , \label{sp}
\end{equation}
where $tr$ denotes the trace, the elements of the scattering phase matrix
are given by
\begin{equation}
{\cal R}_{pp^{\prime}}= \frac{1}{2} tr ({\pmb \sigma}_{p}\, {\bf T}\, 
{\pmb \sigma}_{p^{\prime}}\,
{\bf T}^{\dag}) = \frac{1}{2} \sum_{m m^{\prime}}\, \varphi_{m}\, 
\varphi_{m^{\prime}}^{*}\, {\cal P}_{pp^{\prime}}^{m m^{\prime}}\ ,
\label{scat_mat}
\end{equation}
where the $4 \times 4$ matrix ${\pmb {\cal P}}^{m m^{\prime}}$ is 
the phase matrix for line scattering, which is also called `scattering matrix' 
in the literature.  The product $ \varphi_{m}\, \varphi_{m^{\prime}}^{*}$ 
can be converted into a sum using the conversion formula of 
\cite{stenflo98,bom99}, to which a Doppler convolution is applied. 
In scattering on a two-level atom ($J_{i}=J_{f}=0$), the elements of 
${\pmb {\cal P}}^{m m^{\prime}}$ can be expressed in an elegant form
\begin{equation}
{\cal P}_{pp^{\prime}}^{m m^{\prime}} = {\cal G}_{p}^{m m^{\prime}}
({\pmb k})^{*}\, {\cal G}_{p^{\prime}}^{m m^{\prime}}({\pmb k}^{\prime})\ ,
\label{calp}
\end{equation}
where 
\begin{eqnarray}
{\cal G}_{p}^{m m^{\prime}}({\pmb k})&=&|{\mathcal J}_{L}(\omega)|^{2}
\sum_{\mu \mu^{\prime}} ({\pmb \sigma}_{p})_{\mu \mu^{\prime \prime}}
(-1)^{\mu^{\prime \prime}-m^{\prime}}
(\mu \mu^{\prime \prime})^{h(L)}\nonumber \\ 
&& \times \sum_{l} C(LLl;m -m^{\prime} m_{l})C(LLl;\mu -\mu^{\prime \prime} \mu_{l})
D_{m_{l} \mu_{l}}^{l}(\phi, \theta, 0)\ , 
\label{calg}
\end{eqnarray}
and ${\cal G}_{p^\prime}^{m m^{\prime}}({\pmb k}^\prime)$ is given by a 
similar expression where $p,\,L,\,\omega$ are now replaced by 
$p^\prime,\,L^\prime,\,\omega^\prime$ respectively. Equation~(\ref{scat_mat}) 
can be used to consider scattering in magnetic fields of arbitrary strength 
and orientation. Therefore we refer to ${\pmb {\cal R}}$ as the 
Hanle-Zeeman phase matrix. It is interesting to consider the 
particular case of $0 \to 2 \to 0$ scattering for forbidden transitions
with $L=L^\prime=2$ and study the phase matrix. We consider three domains of 
the field strength $B$, namely \\

\noindent
(i) the case where $B$ is sufficiently strong so that the levels with 
different $m$ are distinct, leading to ``Zeeman scattering'' with $m=m^\prime$ as 
the summation over $m$ in Eq.~(\ref{scat_mat}) drops outs (no $m$-state 
interference)\,; \\
\noindent
(ii) the weak field limit, where the magnetic splitting of the upper level 
is of the same order as the natural width $\Gamma$, so that the quantum 
interferences between magnetic sublevels $m$ and $m^\prime$ take place leading to the 
well-known ``Hanle scattering"\,; and \\
\noindent
(iii) ``Rayleigh scattering" in the absence of magnetic field, when 
the levels with different $m$ are degenerate, so that there is only 
a single phase matrix ${\pmb {\cal P}}^{mm^\prime}\equiv{\pmb {\cal P}}$. \\

\noindent
In the Hanle scattering case, one has to consider totally $25$ pairs of $m$ and 
$m^\prime$ for each element of the $4\times 4$ phase matrix
${\pmb{\cal P}}^{mm^\prime}$ given by Eqs.~(\ref{calp}, \ref{calg}) with 
$L=2$ and $m=m^\prime=2,1,0,-1,-2$, while for the Zeeman scattering, one has 
$5$ phase matrices ${\pmb{\cal P}}^{mm}\equiv {\pmb{\cal P}}^{m}$ with 
$m=\pm 2,\,\pm 1,\,0$. We present these 3 limiting cases  
in the following sections. 

\section{Limiting Cases}
\subsection{Zeeman scattering phase matrix (Rayleigh scattering in strong fields)\label{zee}}
The Zeeman scattering matrix  for $J=0 \to 2 \to 0$ transition in the strong field 
limit can be written as
\begin{equation}
{\pmb{\cal R}}_{Zeeman} = \frac{5}{4}\sum_{m} {\pmb{\cal P}}^{m} H(v_{m},a)\ ,
\end{equation}
where ${\pmb{\cal P}}^{m}$ and $H(v_{m},a)$ represent the scattering phase 
matrices and the Voigt profile functions containing the energies of the upper 
states $|J\, m\rangle$ with $m = -2, \cdots, 2$ respectively. The normalization 
constant is $5/4$.
The frequency dependent shift $v_{m}$ and the damping parameter $a$ 
in Voigt profile function are defined following \cite{stenflo98}.

Zeeman scattering phase matrices ${\pmb{\cal P}}^{m}$ for electric quadrupole 
($E2$) and the magnetic quadrupole ($M2$) forbidden line transitions are 
given below. In the matrix elements, the upper sign represents $M2$ 
transition and the lower sign an $E2$ transition. $M2$ and $E2$ transition 
have the same expression for some matrix elements (eg. the $(1,1)$ element). 

\begin{scriptsize}
\begin{eqnarray}
{\pmb{\cal P}}^{2}=\frac{1}{2}(1-\delta'^{2})(1-\delta^{2})
\left(
\matrix {\frac{1}{2}(1+\delta'^{2})(1+\delta^{2})  &  \pm \frac{1}{2}
(1-\delta'^{2})(1+\delta^{2}) &
 0 & \delta'(1+\delta^{2}) \cr
 \pm \frac{1}{2}(1+\delta'^{2})(1-\delta^{2})&  \frac{1}{2}
 (1-\delta'^{2})(1-\delta^{2}) &
  0 & \pm \delta'(1-\delta^{2}) \cr
  0 & 0 & 0 & 0 \cr
  \delta (1+\delta'^{2}) & \pm \delta (1-\delta'^{2}) & 0 & 2\delta\delta' \cr}
  \right)\ ,
  \end{eqnarray}
      
\begin{eqnarray}
{\pmb{\cal P}}^{-2}=\frac{1}{2}(1-\delta'^{2})(1-\delta^{2})
\left(
\matrix {\frac{1}{2}(1+\delta'^{2})(1+\delta^{2})  &  \pm \frac{1}{2}
(1-\delta'^{2})(1+\delta^{2}) &
 0 & -\delta'(1+\delta^{2}) \cr
 \pm \frac{1}{2}(1+\delta'^{2})(1-\delta^{2})&  \frac{1}{2}
 (1-\delta'^{2})(1-\delta^{2}) &
 0 & \mp \delta'(1-\delta^{2}) \cr
 0 & 0 & 0 & 0 \cr
 -\delta (1+\delta'^{2}) & \mp \delta (1-\delta'^{2}) & 0 & 2\delta\delta' \cr}
\right)\ ,
 \end{eqnarray}

\begin{eqnarray}
{\pmb{\cal P}}^{1}& = \left(
 \matrix { \frac{1}{4}[(1-2\delta'^{2})^{2}+\delta'^{2}]  [(1-2\delta^{2})^{2}
 +\delta^{2}] &
 \mp \frac{1}{4}[(1-2\delta'^{2})^{2}-\delta'^{2}]  [(1-2\delta^{2})^{2}
 +\delta^{2}] & 0 & 0 \cr
 \mp \frac{1}{4} [(1-2\delta'^{2})^{2}+\delta'^{2}]  [(1-2\delta^{2})^{2}-
 \delta^{2}] &
 \frac{1}{4}[(1-2\delta'^{2})^{2}-\delta'^{2} ]  [(1-2\delta^{2})^{2}-
 \delta^{2}] & 0 & 0 \cr
  0 & 0 & 0 & 0 \cr
  -\frac{1}{2}\delta (1-2\delta^{2})[(1-2\delta'^{2})^{2}+\delta'^{2} ] &
 \pm \frac{1}{2}\delta (1-2\delta^{2})[(1-2\delta'^{2})^{2}-\delta'^{2} ]
 & 0 & 0 \cr} \right) & \nonumber \\
 & & \nonumber \\
 & & \nonumber \\
 & +  \left(
 \matrix { 0 & 0 & 0 & - \frac{1}{2}\delta' (1-2\delta'^{2})
 [(1-2\delta^{2})^{2}+\delta^{2} ] \cr
 0 & 0 & 0 & \pm \frac{1}{2}\delta' (1-2\delta'^{2})[(1-2\delta^{2})^{2}
-\delta^{2} ] \cr
0 & 0 & 0 & 0 \cr
0 & 0 & 0 & \delta \delta' (1-2\delta'^{2}) (1-2\delta^{2})\cr }\right)\ ,&
\end{eqnarray}

\begin{eqnarray}
{\pmb{\cal P}}^{-1} &=  \left(
 \matrix { \frac{1}{4}[(1-2\delta'^{2})^{2}+\delta'^{2}]  [(1-2\delta^{2})^{2}
 +\delta^{2}] &
 \mp \frac{1}{4}[(1-2\delta'^{2})^{2}-\delta'^{2}]  [(1-2\delta^{2})^{2}
 +\delta^{2}] & 0 & 0 \cr
 \mp \frac{1}{4} [(1-2\delta'^{2})^{2}+\delta'^{2}]
 [(1-2\delta^{2})^{2}-\delta^{2}] &
 \frac{1}{4}[(1-2\delta'^{2})^{2}-\delta'^{2} ]  [(1-2\delta^{2})^{2}
 -\delta^{2}] & 0 & 0 \cr
  0 & 0 & 0 & 0 \cr
 \frac{1}{2}\delta (1-2\delta^{2})[(1-2\delta'^{2})^{2}+\delta'^{2} ] &
 \mp \frac{1}{2}\delta (1-2\delta^{2})[(1-2\delta'^{2})^{2}-\delta'^{2} ]
 & 0 & 0 \cr} \right)& \nonumber \\
& & \nonumber \\
& & \nonumber \\
& +    \left(
\matrix { 0 & 0 & 0 & \frac{1}{2}\delta' (1-2\delta'^{2})
[(1-2\delta^{2})^{2}+\delta^{2} ] \cr
0 & 0 & 0 & \mp \frac{1}{2}\delta' (1-2\delta'^{2})[(1-2\delta^{2})^{2}
-\delta^{2} ] \cr
0 & 0 & 0 & 0 \cr
0 & 0 & 0 & \delta \delta' (1-2\delta'^{2}) (1-2\delta^{2})\cr }\right)\ ,&
\end{eqnarray}

\begin{eqnarray}
{\pmb{\cal P}}^{0}&=& 9\delta^{2} \delta'^{2}(1-\delta'^{2})(1-\delta^{2})
\left(
\matrix {1 & \mp 1 & 0 & 0 \cr
\mp 1 & 0 & 0 & 0 \cr
0 & 0 & 0 & 0 \cr
0 & 0 & 0 & 0  \cr}
\right)\ ,
\end{eqnarray}
\end{scriptsize}
where the symbols $\delta'=\cos\theta'$ and $\delta=\cos\theta$, with 
$\theta^\prime$ and $\theta$ denoting the angle made by the incoming and 
outgoing ray with respect to the magnetic field chosen to be along the 
$Z$-axis of a coordinate system. 

If we average over all the incoming angles, we obtain the emission 
coefficients $\eta_{I,Q,V}$ used in the thermal radiation limit (LTE). 
$\eta_{U}=0$ for the particular choice of the coordinate system used in this
calculation. However, $\eta_{U}$ can be generated by a clockwise rotation
of the coordinate system through an azimuthal angle $\chi$. We have verified 
that  the new emission 
coefficients $\eta_{I,Q,U,V}$ obtained after rotation are identical with those 
derived by Beckers~\cite{becker} for the quadrupole emission in Zeeman effect 
regime. 

\subsection{Rayleigh scattering phase matrix (the non-magnetic case)
\label{ray}}
The Rayleigh scattering matrix  for $J=0 \to 2 \to 0$ transition in the absence of 
magnetic fields is given by
\begin{equation}
{\pmb{\cal R}}_{Rayleigh}= \frac{5}{4}\, {\pmb{\cal P}}\, H(v,a)\ .
\end{equation}

In this case all the magnetic sublevels of the $J=2$ are degenerate.
Thus the profile functions $\varphi_{m}$ with $m=-2, -1, 0, 1, 2$ become 
identical and equal to $\varphi_{0}$. Therefore we obtain only one scattering 
phase matrix ${\pmb{\cal P}}$, and the profile function $H(v,a)$ corresponding 
to $\varphi_{0}$. The elements of the Rayleigh scattering phase matrix are given by

\begin{scriptsize}
\begin {eqnarray}
{\cal P}_{00}&=& \frac{1}{2}(1-3\delta^{2}+4\delta^{4})
(1-3\delta'^{2}+4\delta'^{4})+ (1-\delta^{2})(1-\delta'^{2})\left[\frac{1}{2}
(1+\delta^{2})(1+\delta'^{2})
+ 9\delta^{2}\delta'^{2}\right]\nonumber \\ & &+
c_{1}(1-\delta^{2})^{1/2}(1-\delta'^{2})^{1/2}\left[4\delta^{3}\delta'^{3}
+6 \delta \delta'(1-2\delta^{2})(1-2\delta'^{2})\right] \nonumber\\
& & + c_{2}\left[6 (1-\delta^{2}) (1-\delta'^{2})\delta^{2}\delta'^{2}
+ \frac{1}{2}(1-5\delta^{2}+ 4\delta^{4})(1-5\delta'^{2}+
4\delta'^{4})\right]\nonumber\\
& & + 4 c_{3}(1-\delta^{2})^{3/2} (1-\delta'^{2})^{3/2}\delta \delta' +
\frac{1}{2}c_{4}(1-\delta^{2})^{2} (1-\delta'^{2})^{2}\ ,
\end{eqnarray}

\begin {eqnarray}
{\cal P}_{01}&=& \mp \frac{1}{2}(1-3\delta^{2}+4\delta^{4})
(1-5\delta'^{2}+4\delta'^{4})+
(1-\delta^{2})(1-\delta'^{2})
\left[\pm \frac{1}{2} (1+\delta^{2})(1-\delta'^{2})
\mp 9\delta^{2}\delta'^{2} \right]\nonumber \\ & &+ c_{1} (1-\delta^{2})^{1/2}(1-\delta'^{2})^{1/2}
\left[ \pm 4 \delta^{3}\delta'(1-\delta'^{2}) \mp
6\delta \delta' (1-2\delta^{2}) (1-2\delta'^{2}) \right] \nonumber \\
& & \mp c_{2} \left[ 6 (1-\delta^{2}) (1-\delta'^{2})\delta^{2}\delta'^{2}
+ \frac{1}{2}(1-5\delta^{2}+ 4\delta^{4}) (1-3\delta'^{2}+ 4\delta'^{4}) \right]
\nonumber \\
& & \pm 4 c_{3} (1-\delta^{2})^{3/2} (1-\delta'^{2})^{1/2}
\delta \delta'^{3} \pm c_{4}\frac{1}{2} (1-\delta^{2})^{2} (1-\delta'^{2})(1+\delta'^{2})\ ,
\end{eqnarray}

\begin {eqnarray}
{\cal P}_{02}&=& \pm s_{1}(1-\delta^{2})^{1/2}(1-\delta'^{2})^{1/2}
 \left[ 2\delta^{3}(1-\delta'^{2}) + 6\delta \delta'^{2}(1-2\delta^{2})\right]
\nonumber \\
& & + s_{2}\left[ \pm \delta' (1-2\delta')
(1-5\delta^{2}+ 4\delta^{4}) \mp 6(1-\delta^{2})(1-\delta'^{2})\delta^{2}
 \delta' \right] \nonumber \\
 & & \mp 2 s_{3} (1-\delta^{2})^{3/2}
 (1-\delta'^{2})^{1/2} \delta (1-3\delta'^{2})
 \pm s_{4}(1-\delta^{2})^{2}(1-\delta'^{2}) \delta' \ ,
\end{eqnarray}

\begin {eqnarray}
{\cal P}_{10}&=&\mp \frac{1}{2}(1-3\delta^{2}+4\delta^{4})(1-5\delta'^{2}+4\delta'^{4})
+(1-\delta^{2})(1-\delta'^{2})\left[\pm \frac{1}{2} (1-\delta^{2})(1+\delta'^{2})
 \mp 9\delta^{2}\delta'^{2} \right]
 \nonumber \\ & &
+ c_{1}(1-\delta^{2})^{1/2} (1-\delta'^{2})^{1/2}\left[
 \pm 4 \delta \delta'^{3} (1-\delta^{2}) \mp
 6\delta \delta' (1-2\delta^{2})  
 (1-2\delta'^{2})\right]
 \nonumber \\ & &
\mp  c_{2}\left[ 6(1-\delta^{2}) (1-\delta'^{2})
 \delta^{2}\delta'^{2} +
 \frac{1}{2}(1-3\delta^{2}+ 4\delta^{4})(1-5\delta'^{2}+ 4\delta'^{4})\right]
 \nonumber \\
 & &\pm 4 c_{3}(1-\delta^{2})^{1/2} (1-\delta'^{2})^{3/2}\delta^{3} \delta'
 \pm \frac{1}{2} c_{4}(1-\delta^{2})(1-\delta'^{2})^{2}(1+\delta^{2})\ ,
 \end{eqnarray}
 
\begin {eqnarray}
{\cal P}_{11}&=&\frac{1}{2}(1-5\delta^{2}+4\delta^{4})(1-5\delta'^{2}+4\delta'^{4})+
(1-\delta^{2})(1-\delta'^{2})\left[ \frac{1}{2} (1-\delta^{2})(1-\delta'^{2})
+9\delta^{2}\delta'^{2} \right] \nonumber \\
& & + c_{1}(1-\delta^{2})^{1/2}
(1-\delta'^{2})^{1/2}\left[ 4 \delta \delta' (1-\delta^{2}) (1-\delta'^{2}) +
6\delta \delta' (1-2\delta^{2}) (1-2\delta'^{2})\right] \nonumber \\
& & + c_{2}\left[ 6(1-\delta^{2}) (1-\delta'^{2})\delta^{2}\delta'^{2} +
\frac{1}{2}(1-3\delta^{2}+ 4\delta^{4})(1-3\delta'^{2}+ 4\delta'^{4})\right]
\nonumber \\
& &+ 4c_{3}(1-\delta^{2})^{1/2} (1-\delta'^{2})^{1/2} \delta^{3} \delta'^{3}
+ \frac{1}{2}c_{4} (1-\delta^{2}) (1-\delta'^{2})
(1+\delta^{2}) (1+\delta'^{2})\ ,
\end{eqnarray}

\begin {eqnarray}
{\cal P}_{12}&=& s_{1}(1-\delta^{2})^{1/2} (1-\delta'^{2})^{1/2}
\left[ 2\delta (1-\delta^{2})(1-\delta'^{2})-6\delta \delta'^{2}
(1-2\delta^{2})\right] \nonumber \\
& &- s_{2}\left[(1-3\delta^{2}+ 4\delta^{4})\delta' (1-2\delta'^{2})-
6(1-\delta^{2})(1-\delta'^{2})\delta^{2}\delta' \right] \nonumber \\
& &-2 s_{3}(1-\delta^{2})^{1/2}
(1-\delta'^{2})^{1/2} \delta^{3} (1-3\delta'^{2})
+ s_{4}(1-\delta^{2})
(1-\delta'^{2}) \delta' (1+\delta^{2})\ , 
\end{eqnarray}

\begin {eqnarray}
{\cal P}_{20}&=&  \mp s_{1}(1-\delta^{2})^{1/2} (1-\delta'^{2})^{1/2}
\left[2(1-\delta^{2})\delta'^{3}+ 6\delta^{2}\delta'(1-2\delta'^{2})\right]\nonumber \\
& & +s_{2}\left[\pm 6(1-\delta^{2}) (1-\delta'^{2})\delta \delta'^{2} \mp
\delta (1-2\delta^{2})(1-5\delta'^{2}+ 4\delta'^{4})\right]\nonumber \\
& & \pm 2 s_{3} (1-\delta^{2})^{1/2}
(1-\delta'^{2})^{3/2} \delta' (1-3\delta^{2})
\mp s_{4} (1-\delta^{2})(1-\delta'^{2})^{2} \delta\ , 
\end{eqnarray}

\begin {eqnarray}
{\cal P}_{21}&=& - s_{1}(1-\delta^{2})^{1/2} (1-\delta'^{2})^{1/2}
\left[ 2\delta'(1-\delta'^{2})(1-\delta^{2})-6\delta^{2}\delta'(1-2\delta'^{2})\right]
\nonumber \\
& & + s_{2}\left[\delta (1-2\delta^{2})(1-3\delta'^{2}+ 4\delta'^{4})
-6(1-\delta^{2}) (1-\delta'^{2})\delta \delta'^{2}\right] \nonumber \\
& & + 2 s_{3}(1-\delta^{2})^{1/2}
(1-\delta'^{2})^{1/2} \delta'^{3} (1-3\delta^{2})
 -s_{4}(1-\delta^{2})(1-\delta'^{2}) \delta (1+\delta'^{2})\ ,
\end{eqnarray}

\begin {eqnarray}
{\cal P}_{22}&=& c_{1}(1-\delta^{2})^{1/2} (1-\delta'^{2})^{1/2}
\left[ (1-\delta^{2}) (1-\delta'^{2})+ 6 \delta^{2}\delta'^{2} \right]\nonumber \\
& & +
c_{2}\left[ 6(1-\delta^{2}) (1-\delta'^{2})\delta \delta'
 +2\delta \delta' (1-2\delta^{2}) (1-2\delta'^{2}) \right] \nonumber \\
 & & +c_{3}(1-\delta^{2})^{1/2} (1-\delta'^{2})^{1/2}
(1-3\delta^{2})(1-3\delta'^{2})
+ 2 c_{4} (1-\delta^{2}) (1-\delta'^{2}) \delta \delta'\ ,
\end{eqnarray}

\begin {eqnarray}
{\cal P}_{33}&=&2 \delta \delta' (1-\delta^{2})(1-\delta'^{2})+
2 \delta \delta' (1-2\delta^{2})(1-2\delta'^{2}) \nonumber \\
& & +c_{1}(1-\delta^{2})^{1/2}(1-\delta'^{2})^{1/2}
\left[ (1-3\delta^{2})(1-3\delta'^{2}) + 6 \delta^{2} \delta'^{2} \right]\nonumber\\
& & +6 c_{2}(1-\delta^{2})(1-\delta'^{2})\delta \delta'
+ c_{3} (1-\delta^{2})^{1/2} (1-\delta'^{2})^{1/2}
 (1-\delta^{2}) (1-\delta'^{2})\ ,
 \end{eqnarray}

 \begin {equation}
 {\cal P}_{03}= {\cal P}_{30}= {\cal P}_{13}= {\cal P}_{31}= {\cal P}_{23}= 
 {\cal P}_{32} = 0\ .
 \end{equation}
 \end{scriptsize}
In the above equations, the symbols $c_{n}=\cos n(\phi-\phi')$ and 
$s_{n}=\sin n(\phi-\phi')$. The incoming and outgoing rays make an 
angle $(\theta^\prime, \phi^\prime)$ and $(\theta, \phi)$ with respect to 
the $Z$-axis of a coordinate system. 

\subsection{Hanle scattering phase matrix (in weak field limit)
\label{han}}
In the weak-field limit, the magnetic splitting is of the same order as the 
natural line width of an atomic state. Therefore there exists a frequency 
coupling between the magnetic sublevels of the  upper energy level. 
If the product of profile 
functions $\varphi_{m}\varphi^{*}_{m^\prime}$ is simplified as mentioned before 
(see the paragraph following Eq.~(\ref{scat_mat})), the weak field Hanle 
phase matrix is obtained.

\section{Numerical results for a $0 \to 2 \to 0$ transition
\label{num_res}}
We consider the example of single scattering involving a $0 \to 2 \to 0$ 
forbidden transition. An unpolarized beam of radiation is incident 
on the atom. Therefore the four Stokes parameters $(I, Q, U,V)$  of the 
scattered radiation are $({\cal R}_{00}, {\cal R}_{10},{\cal R}_{20}, 
{\cal R}_{30})$ respectively. In the 
expressions~(\ref{for_scat_rayleigh}) - (\ref{h3}) for the Stokes 
parameters, the upper sign represents $M2$ type transition and the 
lower sign represents $E2$ type transition. 
The different curves in all the figures represent different values of
magnetic field strength defined by a splitting parameter 
$v_{B}=g\nu_{\rm L}/\Delta\nu_{\rm D}$, where $\nu_{\rm L}$ is the Larmor 
frequency, $g$ the Land\'e factor, and $\Delta\nu_{\rm D}$ the Doppler 
width. The solid curve corresponds to $v_{B}=0.004$, the dotted curve
to $v_{B}=0.02$, the dot-dashed curve to $0.1$, the dashed curve to $0.5$
 and the long-dashed curve to $2.5$. The damping parameter of the 
Voigt profile function $a=0.004$. 
The direction of the magnetic field is chosen along the $Z$-axis of 
the co-ordinate system. 

\subsection{Forward scattering\label{num_res_for}}
\begin{figure*}[ht]
\vspace*{10mm}
\begin{center}
\includegraphics[width=14.0cm,height=5cm]{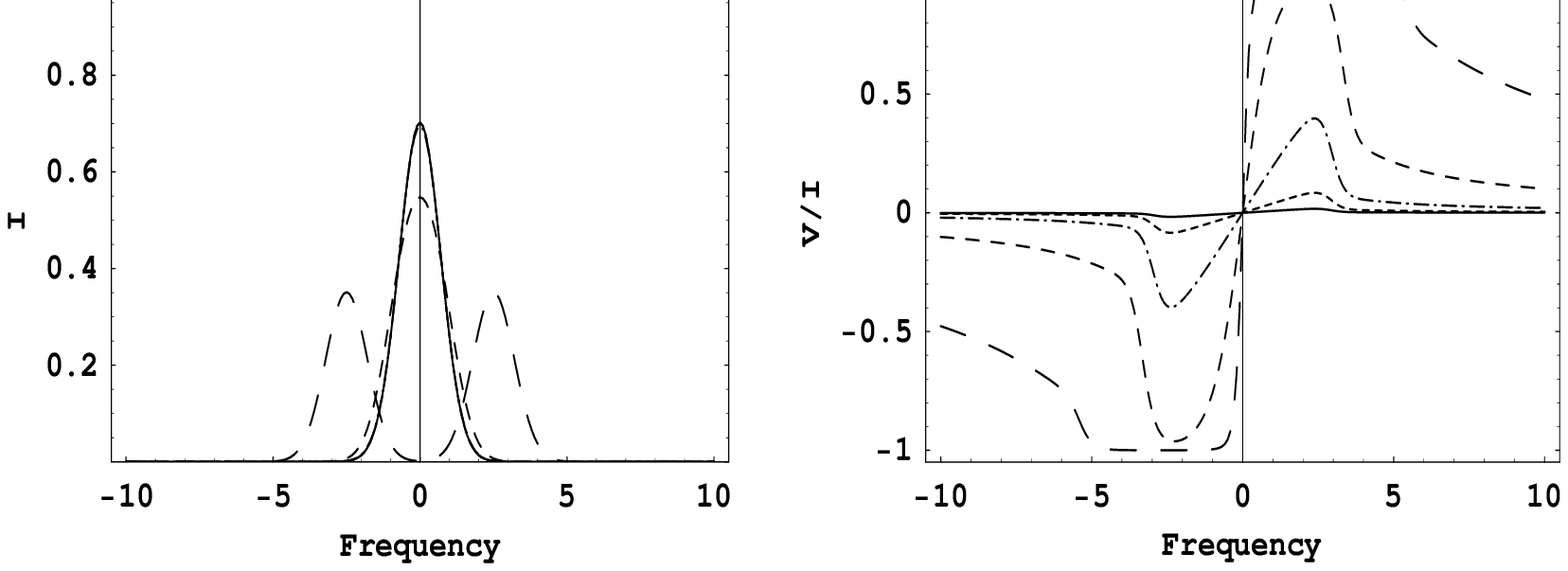}
\end{center}
\caption{The Stokes profiles in the regime of Hanle-Zeeman scattering. 
The case of forward scattering is considered. Different curves represent 
various values of the field strength $v_B$. See sections~\ref{num_res} and 
\ref{num_res_for} for model parameters. 
\label{set1}}
\end{figure*}

Figure~\ref{set1} shows the scattered Stokes profiles in a forward 
scattering event ($\theta'=0^\circ,\, \phi'=0^\circ$; $\theta=0^\circ,\, \phi=0^\circ$). For this choice of scattering geometry, the Stokes 
parameters, given by the following expressions, are the same for both 
$M2$ and $E2$ type transitions. For the case of 
non-magnetic Rayleigh scattering\,: 
\begin{equation}
I = \frac{5}{2}\left[\varphi_{0}\, \varphi_{0}^{*}\right];
\hspace{0.3cm}Q  = 0; \hspace{0.3cm}
U = 0;\hspace{0.3cm} V = 0.
\label{for_scat_rayleigh}
\end{equation}
Likewise, the scattered Stokes profiles for the Hanle-Zeeman scattering 
are\,:  
\begin{equation}
\!\!\!\!\!\!\!\!\!\!\!\!
I = \frac{5}{4}\left[\varphi_{1}\, \varphi_{1}^{*} + \varphi_{-1}\, 
\varphi_{-1}^{*}\right];
\quad Q= 0; \quad 
U = 0; \quad
V = \frac{5}{4}\left[\varphi_{1}\, \varphi_{1}^{*} - \varphi_{-1}\, 
\varphi_{-1}^{*}\right].
\end{equation}

In the particular geometry of forward scattering, the same expressions hold 
good for the Zeeman scattering case also. Therefore we show only $I$ and 
$V/I$ profiles corresponding to the Hanle-Zeeman case. The Stokes $I$ 
for the Rayleigh scattering case is nearly same as the Stokes $I$ of 
the Hanle-Zeeman case (the solid curve). The Stokes $Q/I=0$ in 
both the cases.
It can be understood from the classical theory that the degree 
of linear polarization is zero for the forward
Rayleigh scattering. As the line of sight is taken parallel
to the magnetic field, the standard doublet pattern of longitudinal 
Zeeman effect is seen in $I$. The $V/I$ profiles show anti-symmetric 
pattern (oppositely circularly polarized). These components are produced 
due to the scattering involving $|0,0\rangle$ and $|2,\pm 1\rangle$ 
states. The resolution of $I$ into 2 components takes place only for 
very strong fields like $v_B=2.5$. For weaker fields, it appears only 
as a slight broadening of the non-magnetic $I$ profiles. All the 
weak field cases for $v_B \le 0.1$ merge, and are not distinguishable 
graphically. 

\subsection{Scattering at $90^\circ$}
This scattering geometry corresponds to the maximum degree of linear 
polarization. The angles chosen for this case are\,: 
$\theta'=0^\circ,\, \phi'=0^\circ$ and $\theta=90^\circ,\,\phi=45^\circ$.
In this case, the scattered Stokes profiles are given by 
\begin{equation}
I = \frac{5}{4}\left[\varphi_0\, \varphi^{*}_0\right]; \hspace{0.3cm}
Q  = \mp \frac{5}{4}\left[\varphi_0\, \varphi^{*}_0\right]; \hspace{0.3cm}
U = 0;\hspace{0.3cm} V = 0, \label{eq33}
\end{equation}
for Rayleigh scattering case, and 
\begin{equation}
\!\!\!\!\!\!\!\!\!\!\!\!\!\!\!\!
I = \frac{5}{8}\left[\varphi_{1}\, \varphi_{1}^{*} + \varphi_{-1}\, 
\varphi_{-1}^{*}\right]; \quad 
Q = \mp \frac{5}{8}\left[\varphi_{1}\, \varphi_{1}^{*} + \varphi_{-1}\, 
\varphi_{-1}^{*}\right]; \quad 
U =0; \quad
V = 0, \label{eq34}
\end{equation}
for the Hanle-Zeeman as well as the Zeeman scattering cases. 

It is clearly seen that the maximum degree of linear polarization 
$Q/I=-1$ in all the three cases. The $Q/I$ 
ratio is independent of frequency as well as the magnetic field strength. 
From Eqs.~(\ref{eq33}) and (\ref{eq34}), we observe that the 
linear polarization ($Q$) profiles for $M2$ and $E2$ type 
transitions differ only in sign. As the $I$ profiles look very similar to the 
$I$ profiles of Fig.~\ref{set1}, except for the fact that $U/I=V/I=0$ and 
$Q/I=-1$, we do not show these profiles again. 

\subsection{Scattering at an arbitrary angle\label{num_res_arb}}
\begin{figure*}[ht]
\vspace*{10mm}
\begin{center}
\includegraphics[width=14.0cm,height=14.0cm]{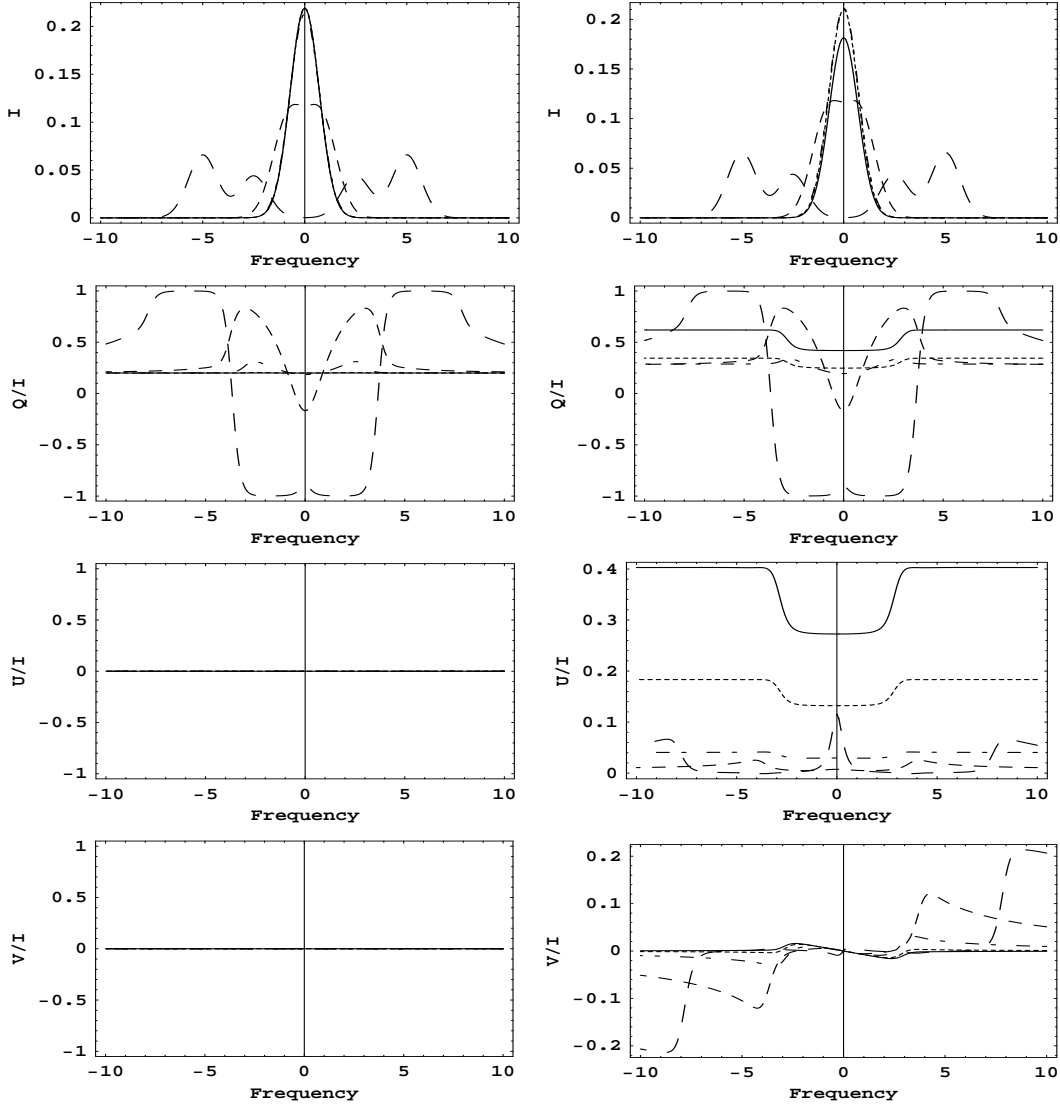}
\end{center}
\caption{Zeeman
($1^{\rm st}$ column) and Hanle-Zeeman scattering ($2^{\rm nd}$ column) in
a $M2$ type scattering event. The scattering geometry is defined through the 
choice of angles\,: $\theta^{\prime}=45^\circ, \phi^{\prime}=0^\circ$
and $\theta=90^\circ, \phi=45^\circ$. Different curves correspond to different 
values of the field strengths expressed through the parameter $v_B$. 
The solid curves: $v_{B}=0.004$, dotted curves:
$0.02$, dot-dashed curves: $0.1$, dashed curves: $0.5$, and long
dashed curves: $2.5$. See section~\ref{num_res_arb} for discussions. \label{set3}}
\end{figure*}

\begin{figure*}[ht]
\vspace*{10mm}
\begin{center}
\includegraphics[width=14.0cm,height=14.0cm]{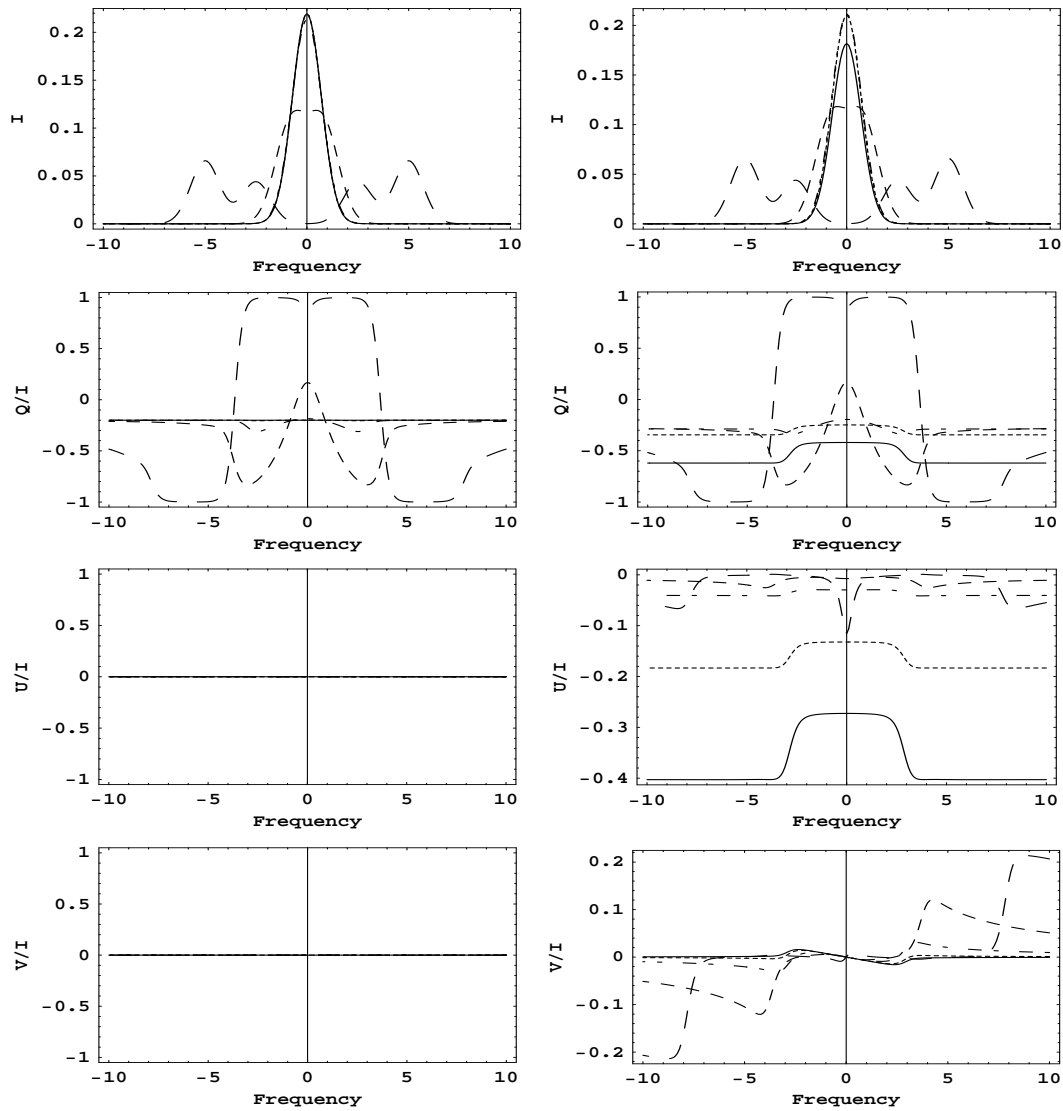}
\end{center}
\caption{Zeeman
($1^{\rm st}$ column) and Hanle-Zeeman scattering ($2^{\rm nd}$ column) in
a $E2$ type single scattering event. 
The scattering geometry and different curve types are the same as in 
Fig.~\ref{set3}. \label{eset3}}
\end{figure*}

In Figs.~\ref{set3} and \ref{eset3} we show the results for $M2$ and 
$E2$ type transitions respectively. They 
correspond to the choice of scattering geometry\,: $\theta'=45^\circ,\, 
\phi'=0^\circ$ and $\theta=90^\circ,\, \phi=45^\circ$.  
For this specific choice of angles, the scattered Stokes parameters are
given by the following simple expressions\,:
\begin{equation}
\label{r3}
I = \frac{5}{8}\,\left[\varphi_0 \varphi^{\ast}_0\right]\ ; \hspace{0.3cm}
Q  = 0\ ; \hspace{0.3cm}
U = 0\ ;\hspace{0.3cm} V = 0\ ,
\end{equation}
for non-magnetic Rayleigh scattering case. For the Zeeman scattering 
case we have 
\begin{eqnarray}
\label{z3}
I &=& \frac{5}{4}\,\left[0.125\,(\varphi_{1} \varphi_{1}^{\ast} + 
\varphi_{-1} \varphi_{-1}^{\ast})+
0.1875\,(\varphi_{2} \varphi_{2}^{\ast} + \varphi_{-2} 
\varphi_{-2}^{\ast})\right], \nonumber \\
Q &=& \frac{5}{4}\,\left[\mp 0.125\,(\varphi_{1} \varphi_{1}^{\ast} + 
\varphi_{-1} \varphi_{-1}^{\ast}) \pm
0.1875\,(\varphi_{2} \varphi_{2}^{\ast} + \varphi_{-2} 
\varphi_{-2}^{\ast})\right],
\end{eqnarray}
with $U=V=0$, and finally for the Hanle-Zeeman case we get 
\begin{eqnarray}
\label{h3}
I &=& \frac{5}{4}\,\big[0.125\,(\varphi_{1} \varphi_{1}^{\ast} + 
\varphi_{-1} \varphi_{-1}^{\ast})+
0.1875\,(\varphi_{2} \varphi_{2}^{\ast} + \varphi_{-2} 
\varphi_{-2}^{\ast}) \nonumber \\
& & + {\rm i}\times 0.125\,(\varphi_{-1} \varphi_{1}^{\ast} - 
\varphi_{1} \varphi_{-1}^{\ast})-
0.0625\,(\varphi_{2} \varphi_{-2}^{\ast} + \varphi_{-2} \varphi_{2}^{\ast})\big], \nonumber \\
Q &=& \frac{5}{4}\,\big[\mp 0.125\,(\varphi_{1} \varphi_{1}^{\ast} + 
\varphi_{-1} \varphi_{-1}^{\ast})\pm
0.1875\,(\varphi_{2} \varphi_{2}^{\ast} + \varphi_{-2} 
\varphi_{-2}^{\ast})\nonumber \\
&&\mp {\rm i}\times 0.125\,(\varphi_{-1} \varphi_{1}^{\ast} - \varphi_{1} 
\varphi_{-1}^{\ast})\mp
0.0625\,(\varphi_{2} \varphi_{-2}^{\ast} + \varphi_{-2} \varphi_{2}^{\ast})
\big], \nonumber \\
U &=&\frac{5}{4}\times 0.0883883\,\big[(1-{\rm i})\,(\pm \varphi_{-2} 
\varphi_{1}^{\ast} 
\mp \varphi_{2} \varphi_{1}^{\ast} \pm \varphi_{-1} \varphi_{2}^{\ast}\mp
\varphi_{-1} \varphi_{-2}^{\ast})\nonumber \\ && 
+ (1+{\rm i})\,(\pm \varphi_{1} \varphi_{-2}^{\ast}\mp
\varphi_{1} \varphi_{2}^{\ast}\pm \varphi_{2} \varphi_{-1}^{\ast}\mp
\varphi_{-2} \varphi_{-1}^{\ast})
\big],\nonumber \\
V &=& \frac{5}{4}\times 0.0883883\,\big[(1-{\rm i})\,(\varphi_{1} 
\varphi_{-2}^{\ast}
-\varphi_{1} \varphi_{2}^{\ast}+\varphi_{-2} \varphi_{-1}^{\ast}-\varphi_{2} 
\varphi_{-1}^{\ast}) 
\nonumber \\ && + (1+{\rm i})\,(\varphi_{-2} \varphi_{1}^{\ast}-
\varphi_{2} \varphi_{1}^{\ast}+\varphi_{-1} \varphi_{-2}^{\ast}-
\varphi_{-1} \varphi_{2}^{\ast})
\big]. 
\end{eqnarray}
Clearly the $Q/I$ and $U/I$ in Figs.~\ref{set3} and \ref{eset3} for both 
Zeeman as well as Hanle-Zeeman cases differ only in sign (see Eqs.~(\ref{z3})
and (\ref{h3})). However the $V/I$ remains the same for both $M2$ and $E2$ 
type transitions. In the following we discuss only Fig.~\ref{set3} for $M2$ 
type transition.

In the cases of Zeeman and Hanle-Zeeman scattering, the Stokes $I$ profiles  
clearly show splitting in the strong fields. When the fields 
are weak, these components are not resolved. In this choice of geometry, 
the Stokes $Q$ gets contributions not only from the energy 
states $|2,\pm 1\rangle$  but also from  $|2,\pm 2\rangle$ in both 
the cases (see Eqs.~(\ref{z3}) and (\ref{h3})). Rayleigh scattered Stokes 
$Q$ is zero in the $M2$ as well as $E2$ transition, even for a non-zero 
scattering angle (see Eq.~(\ref{r3})). The $I$ profile for Rayleigh scattering 
is similar to the solid curve of the general Hanle-Zeeman case, and hence is 
not shown again.  
Equations~(\ref{z3}) and (\ref{h3}), differ only in the $m$-state 
interference terms. 
The effect of these extra terms in Eq.~(\ref{h3}) 
becomes significant only 
for weak fields (compare solid, dotted and dot-dashed $Q/I$ curves in the 
2nd row of Fig.~\ref{set3}). When the fields are stronger 
(dashed and long-dashed curves), the Zeeman and 
Hanle-Zeeman scattering give nearly same  
$Q/I$ profiles, showing that the interference terms 
gradually become negligible. 

The Stokes $U$ and $V$ are zero in Rayleigh and Zeeman scattering cases due to 
our choice of the scattering geometry, but they appear in Hanle-Zeeman
scattering cases in the weak field limit. The generation of these $U$ and $V$ 
in the Hanle-Zeeman scattering case is entirely due to the interference 
between different $m$-states. As the magnetic field strength increases, $U/I$ 
decreases and approaches small values. The $V/I$ is extremely small in the 
weak field limit and gradually develops the typical anti-symmetric profiles of 
the usual Zeeman effect as the field strength increases.

\section{Conclusions}
The polarization phase matrices for forbidden line transitions 
are derived. The 3 important regimes
of Rayleigh, Zeeman, and Hanle-Zeeman scattering are studied.
The regime of Hanle-Zeeman scattering 
provides a complete description of scattering theory 
in magnetic fields of arbitrary strengths, with the other two regimes 
serving as extreme limiting cases. We show that the linear polarization 
profiles ($Q$ and $U$) for $M2$ and 
$E2$ type transitions differ only in sign. The $V$ profiles 
are however, insensitive to the type of transition.

\ack 
YYO and GR thank Mr. H. S. Nataraj for useful discussions.
Part of this work was done in the Department of Physics, Osaka University.
YYO likes to thank the Department of
Physics, Osaka University for extending the research facilities. She also
acknowledges the Matsumae International Foundation, Japan for financial 
support to visit Japan.

\newpage

\newpage

\newpage
\end{document}